\documentclass[twocolumn]{svjour3}          

\smartqed  
\usepackage{graphicx}
\usepackage{textcomp}
\usepackage{amsmath}
\usepackage{amssymb}
\usepackage{hyphenat}
\usepackage{booktabs}
\usepackage{hyperref}
\hypersetup{hidelinks,colorlinks = true,linkcolor = blue,urlcolor  = blue,citecolor = blue,anchorcolor = blue}

\usepackage{doi}
\usepackage[square,numbers,sort&compress]{natbib}

\journalname{Theor Chem Acc}

\begin{document}

\title{Enhancing sampling in atomistic simulations of solid state materials for batteries: a focus on olivine NaFePO$_4$
}

	\author{Bruno Escribano \and Ariel Lozano \and  Tijana Radivojevi\'{c} \and Mario~Fern\'{a}ndez-Pend\'{a}s \and
	        Javier Carrasco \and Elena Akhmatskaya
}

\institute{B. Escribano \and T. Radivojevi\'{c} \and M. Fern\'{a}ndez-Pend\'{a}s \at
              Basque Center for Applied Mathematics, Alameda de Mazarredo 14 (48009) Bilbao, Bizkaia, Spain \\
              \email{bescribano@bcamath.org}           
           \and
           A. Lozano \at
              Basque Center for Applied Mathematics, Alameda de Mazarredo 14 (48009) Bilbao, Bizkaia, Spain \\
              CIC EnergiGUNE, Albert Einstein 48 (01510) Mi\~{n}ano, \'{A}lava, Spain \\
              \email{alozano@bcamath.org}
           \and 
           J. Carrasco \at
              CIC EnergiGUNE, Albert Einstein 48 (01510) Mi\~{n}ano, \'{A}lava, Spain \\
           \and
           E. Akhmatskaya  \at
              Basque Center for Applied Mathematics, Alameda de Mazarredo 14 (48009) Bilbao, Bizkaia, Spain \\          
              IKERBASQUE, Basque Foundation for Science, E-48013 Bilbao, Spain \\
}

\date{Received: date / Accepted: date}

\maketitle

\begin{abstract}

The study of ion transport in electrochemically active materials for energy storage systems requires
simulations on quantum- atomistic- and meso-scales.
The methods accessing these scales not only have to be effective but also well compatible to provide a full description of the underlying processes.
We propose to adapt the Generalized Shadow Hybrid Monte Carlo (GSHMC) method to atomistic simulation of ion intercalation electrode materials for batteries.
The method has never been applied to simulations in solid state chemistry but it has been successfully used for simulation of biological macromolecules, demonstrating better performance and accuracy than can be achieved with the popular molecular dynamics (MD) method.
It has been also extended to simulations on meso-scales, making it even more attractive for simulation of battery materials.
We combine GSHMC with the dynamical Core-Shell model to incorporate polarizability into the simulation and apply the new Modified Adaptive Integration Approach, MAIA, which allows for a larger time step due to its excellent conservation properties. 
Also, we modify the GSHMC method, without losing its performance and accuracy, to reduce the negative effect of introducing a shell mass within a dynamical shell model. 
The proposed approach has been tested on olivine NaFePO$_4$, which is a promising cathode material for Na-ion batteries.
The calculated Na-ion diffusion and structural properties have been compared with the available experimental data and with the results obtained using MD and the original GSHMC method.
Based on these tests, we claim that the new technique is advantageous over MD and the conventional GSHMC and can be recommended for studies of other solid-state electrode and electrolyte materials whenever high accuracy and efficient sampling are critical for obtaining tractable simulation results.     


\end{abstract}

\section{Introduction}

The development of advanced materials for energy storage has grown into a topic of intense research due to their importance in powering portable devices, electric vehicles, and  electrical grids collecting energy from renewable sources.
During the last decade, Li-ion rechargeable batteries have become a gold standard in storing electrical energy  \cite{Tarascon2001,Etacheri2011,Armand2008,Goodenough2013}.
However, in an ever-growing demand for better batteries, low cost and natural abundance of precursor materials are quickly emerging as the basis for \textit{beyond Li-ion} technology. 
In this context, the fifth most abundant element in the earth crust and the second lightest and smallest alkali metal after Li, Na, is currently considered as a natural candidate for the next generation of low cost batteries \cite{Yabuuchi2014,Han2015}.

Therefore, research on active materials for Na-based technologies is gathering momentum \cite{Choi2016}. 
Atomic-scale computational approaches are becoming increasingly useful for these exploratory studies in order to avoid time-consuming trial and error approaches \cite{Hautier2012,Jain2016}. 
Based on ab initio methods and modern information technology tools, these techniques enable the assessment of critical properties of interest such as phase stability, electronic structure, and ionic conductivity \cite{Islam2014,Meng2009,Meng2013,Ramzan2009,Moreau2010,Zhou2004a,Dathar2011,Wang2007}.
However, the computation of some of these properties can be very time consuming and, therefore, impracticable to tackle from a pure ab initio viewpoint.
This is particularly true for solid state ionics and ion intercalation processes in electrolyte and electrode materials for batteries, where one should deal with the mobility of alkali ions.

Typically, ion diffusion occurs over long timescales and its statistically meaningful study usually requires to model systems containing thousands of atoms at least. 
Such simulations are not currently feasible with the use of ab initio methods.   
Classical interatomic potentials (force fields) are a practical solution to this problem in many cases, since such methods reduce the electronic degrees of freedom and thus allow for handling longer timescales and larger system sizes.
Many studies of electrolyte and electrode materials based on interatomic potentials have been performed in the last two decades, mainly dealing with statical energy evaluations to determine ion diffusion paths and activation energies, defect chemistry, and stability of surfaces and nanostructures \cite{Islam2014}.

In spite of such success, classical interatomic potentials are often inefficient to properly account for rare events, specially when they are applied in molecular dynamics (MD) simulations.
Ion diffusion processes are indeed rare events that involve ion hopping between adjacent sites and, sometimes, even colective ion transport. 
In order to properly simulate such phenomena in technologically relevant materials, very long simulations are normally required (see, e.g., Ref. \cite{Burbano2016}). 
To overcome this issue, different proposals for enhancing sampling efficiency in MD simulations of ion diffusion have been reported.
For example, one can modify the particles momenta on the fly to stimulate the events of interest (ion particles jumps), but these methods do not preserve the desired distribution \cite{Boulfelfel2011}. 
A similar approach is to rely on very high unphysical temperatures to force the observation of rare diffusion events \cite{Tealdi2012,Yang2011}.
The so-called Generalized Shadow Hybrid Monte Carlo method (GSHMC) is another promising technique, which has been proven to be successful when applied to the study of rare events in complex biological processes \cite{Wee2008,Akhmatskaya2011,Escribano2015}, but it has not been used for computing properties of solid crystalline systems yet.

In this work we investigate the effectiveness of enhanced sampling approaches in the simulation of various properties of olivine NaFePO$_4$ using GSHMC-based techniques.
We focused on NaFePO$_4$ because this system is a promising candidate as a cathode material for Na-ion batteries \cite{Galceran2014b}.
It is the Na counterpart of LiFePO$_4$, which is used in many commercial Li-ion batteries nowadays \cite{Yuan2011}.
In contrast to the Li case, NaFePO$_4$, forms a stable partially sodiated structure Na$_{2/3}$FePO$_4$ upon charge \cite{Galceran2014b,Zhu2013}  or chemical Na intercalation \cite{Saracibar2016,Ramzan2009}.
The NaFePO$_4$ and Na$_{2/3}$FePO$_4$ systems offer us the opportunity to test the GSHMC sampling approach in a technologically relevant material and they are complex enough to analyze the performance of different sampling techniques.
In addition, a force field specifically developed for olivine NaFePO$_4$ already exists \cite{Whiteside2014}.

The paper is organized as follows. 
In Section \ref{sec:interatomic} we describe the force field used to model the bulk NaFePO$_4$. 
Then, in Section \ref{sec:sampling} we summarize the basics of the GSHMC method and explain the additional modifications that we have introduced to the original proposal. 
Section \ref{sec:numerical} compares the efficiency in terms of accuracy and performance of two variants of GSHMC and the standard MD method to account for  structural and dynamical properties of the bulk NaFePO$_4$ and Na$_{2/3}$FePO$_4$. 
Finally, conclusions are presented in Section \ref{sec:conclu}.

\section{Computational model}
\label{sec:interatomic}

The force field proposed for olivine NaFePO$_4$ by Whiteside et al. \cite{Whiteside2014} follows the Born model, with the addition of shells to some ions. The shell-model is introduced to describe the ionic polarization as suggested by Dick and Overhauser \cite{Dick1958}. 
In this model an ion is described using a central core with a charge $X$ and a shell of a charge $Y$. 
These two charges are balanced so that the sum of $(X + Y)$ is the same as the valence state of the ion. 
A core and a shell are coupled together in a \textit{core-shell unit} via a harmonic potential, which allows the shell to move with respect to the core, thus simulating a dielectric polarization.

The total potential energy is given by
\begin{equation}
U =  V_\text{C} + V_\text{BH} + V_\text{CS},
\label{eq:fullpotential} 
\end{equation} 
where $V_\text{C}$ stands for the long-range Coulomb interactions, $V_\text{BH}$ is a Buckingham potential that models short-range repulsions and van der Waals forces between atoms, and $V_\text{CS}$ is the interaction within each core-shell unit. In Eq. \eqref{eq:fullpotential}, the Coulomb interactions are computed between every pair of charged particles in the system but not within a core-shell unit. The short-range potential is considered solely between shells when core-shell units are involved and $V_\text{CS}$ is computed for each core-shell unit.

The terms in Eq. \eqref{eq:fullpotential} are explicitly given by
\begin{equation}
\nonumber  
V_\text{C}(r_{ij}) = \frac{1}{4\pi\epsilon_0}\sum_{i,j=1}^N\frac{q_i q_j}{r_{ij}},
\label{eq:coulomb}
\end{equation} 
where $\epsilon_0$ is the vacuum permitivity, $r_{ij}$ is the distance between particles $i$ and $j$, $q_i$ and $q_j$ are their respective charges and $N$ is the number of particles,
\begin{equation} 
\nonumber
V_\text{BH}(r_{ij}) = \sum_{i,j=1}^N A_{ij} \exp \left(-\frac{r_{ij}}{\rho_{ij}} \right) - \frac{C_{ij}}{r_{ij}^6},
\label{eq:buckingham} 
\end{equation} 
where $A_{ij}$, $\rho_{ij}$, and $C_{ij}$ are positive constants defining the shapes of the repulsive and the attractive terms of the potential, and
\begin{equation} 
\nonumber
V_\text{CS}(r_{l}) = \sum_{l=1}^L \frac{1}{2} \ k_l \ r_{l}^2,
\label{eq:coreshell} 
\end{equation} 
where $k_l$ is the spring constant for the $l$-th core-shell unit, $r_{l}$ is the displacement between the shell center and its core, and $L$ is the total number of shells. 

In the work by Whiteside et al. \cite{Whiteside2014} an extra three-body bonding term for the O-P-O angles in the PO$_4$ tetrahedral units was also included. It takes the form of a harmonic angle-bending potential given by 
\begin{equation} 
	\nonumber
V_\text{Ang}(\theta_k) = \sum_{k=1}^K \frac{1}{2} k_\text{ang} (\theta_k - \theta_0)^2,
\label{eq:angular} 
\end{equation} 
where $k_\text{ang}$ is the spring constant, $\theta_0$ is the equilibrium bond angle, $\theta_k$ is the current value of the bond $k$, and $K$ is the total number of angle interactions. 

For this study, we took from Ref. \cite{Whiteside2014} the full set of parameters defining the force field for olivine NaFePO$_4$ (Table \ref{tab:forcefield}).

\begin{table}
        \centering
		\caption{Force field parameters for olivine NaFePO$_4$ taken from Ref.~\cite{Whiteside2014}.} 
		\begin{tabular}{ c c c c  }
			BH & & &  \\ \toprule
			Interaction                   &  $A$ (eV)    & $\rho$ (\AA)  & $C$ (eV \AA$^6$)    \\ \midrule
			Na$^+$ - O$_\text{sh}^{2-}$            &  629.757635  &  0.317034    &  0.0   \\[3pt] 
			Fe$_\text{sh}^{2+}$- O$_\text{sh}^{2-}$ &  1105.2409   &  0.3106     &  0.0   \\[3pt] 
			P$^{5+}$ - O$_\text{sh}^{2-}$             &  897.2648    &  0.3577   &  0.0   \\[3pt] 
			O$_\text{sh}^{2-}$ - O$_\text{sh}^{2-}$ &  22764.3     &  0.149      & 44.53  \\
			& & &  \\
			CS & & & \\ \toprule
			Species  & Core charge     & Shell charge & $k$ (eV \AA$^{-2}$)       \\ \midrule
			Fe$^{2+}$       & -0.997          & 2.997        & 19.26  \\ 
			O$^{2-}$        &  0.96           & -2.96        & 65.0   \\
			&&& \\
		\end{tabular}
		\begin{tabular}{ c c c }
			Ang& & \\ \toprule    
			bond     &     $k_\text{ang}$ (eV rad$^{-2}$)      &   $\theta_0$ (deg)  \\ \midrule
			O$_\text{sh}^{2-}$-P$^{5+}$-O$_\text{sh}^{2-}$    &  1.322626  &  109.47 \\
		\end{tabular}
		\label{tab:forcefield}
\end{table}

At this point, it must be mentioned an important issue regarding molecular dynamics simulations based on a Core-Shell potential model. 
In the original Core-Shell model, shell particles are massless and the model requires them to be always at their optimal positions with zero forces \cite{Dick1958}.
When atomic motions are considered during dynamical simulations the shells should respond instantaneously to the motions of the cores.
Two main approaches are found in the literature to deal with the integration of equations of motion in this case: the so-called \textit{shell relaxation} (CS-min) scheme \cite{Lindan1993} and the \textit{adiabatic shells} (CS-adi) method \cite{Mitchell1993}. 

The CS-min approach consists of three steps: i) to calculate the forces on all cores with the shells fully relaxed; ii) to update the core positions using the forces; iii) to relax the shells for the new core positions \cite{Lindan1993}. 
The last step involves the energy minimization in the multidimensional space of shell configurations which turns to be a very computationally demanding task. 
The CS-adi scheme was proposed as a faster alternative to the CS-min method. 
In the CS-adi approach, a small fraction $x$ of the ion mass is put on the shell, whereas the remaining (1-$x$) fraction belongs to the core. 
Then, all the particles positions propagate following the conventional MD technique \cite{Mitchell1993}. 
Having sufficiently small masses, the shells adiabatically follow the cores motion during the simulation. 
A proper choice of the mass distribution for the core-shell units is crucial for the accuracy of the method. Care has to be taken to ensure the negligible effect of an extra thermal energy, introduced by the relative motion between a core and its shell, on the kinetic energy of the simulated system.
However, to date there is no a systematic way to assign mass values for shells. 

In this study we choose to apply the adiabatic shell scheme due to its computational efficiency, and propose a novel approach for introducing a shell mass in the way that reduces its negative effect on the kinetic energy of the system.

\section{Sampling}
\label{sec:sampling}

Our choice of the simulation technique for modeling olivine NaFePO$_4$ has been based on two requirements. 
We looked for an enhanced sampling method, which can efficiently sample multi-dimensional space and detect the rare events, as well as be easily extended for simulations on meso-scales. Such properties are critical for effective study of ion transport in bulk and nanostructured materials.

The Generalized Shadow Hybrid Monte Carlo me-thod or GSHMC by Akhmatskaya and Reich was originally developed for efficient atomistic simulation of complex systems \cite{Akhmatskaya2008} and then adjusted to simulation on meso-scales, without losing its capacity for exact sampling at the target temperature \cite{Akhmatskaya2011}. 
The method however has never been applied to solid state chemistry. 
In this study we investigate the performance of GSHMC in simulation of olivine NaFePO$_4$ and propose some modifications to the original algorithm aiming to improve its accuracy and sampling efficiency specifically  in simulation of battery materials. 

 \subsection{GSHMC: Generalized Shadow Hybrid Monte Carlo}

The GSHMC method is a type of Markov Chain Monte Carlo, with better sampling performance than Monte Carlo or MD in molecular simulations and with a negligible computing overhead. 
GSHMC is especially appropriate when exploring configurational spaces of high dimensionality, finding global energy minima, and simulating rare events such as phase transitions.
Its theoretical foundation has already been published elsewhere \cite{Akhmatskaya2008,Akhmatskaya2012,Escribano2015,AkhmatskayaPatent,Akhmatskaya2011}.
It has recently been implemented in an open-source MD package \cite{Escribano2013,Fernandez2014} and applied to the study of proteins \cite{Mujika2012,Wee2008}. 
In the following lines we present a brief summary of the method.

Essentially, GSHMC is a Hybrid Monte Carlo (HMC) method \cite{Duane1987} that aims to achieve high efficiency by sampling with respect to modified energies (modified or shadow Hamiltonians). 
At the same time it preserves most of the dynamical information by applying a partial momentum update instead of fully resampling the momenta between molecular dynamics trajectories, as is the case of HMC.

Shadow Hamiltonians are asymptotic expansions of the true Hamiltonian in powers of the time step $\Delta t$.
They are conserved better than true Hamiltonians by symplectic integrators such as the leapfrog / Verlet algorithm commonly used in molecular simulations \cite{Hairer2002}.
Thus replacing Hamiltonians with shadow Hamiltonians in Metropolis tests leads to higher acceptance rates than those obtained in the HMC method.   
The computational cost required for the evaluation of shadow Hamiltonians is negligible compared to the force evaluation in an MD simulation. 
Efficient algorithms for computing modified energies can be found for example in Refs. \cite{Skeel2001,Akhmatskaya2008,Radivojevic2,RadivojevicThesis}. 
The GSHMC method employs the Lagrangian formulation of shadow Hamiltonians of an arbitrary order for the leapfrog integrator \cite{Akhmatskaya2008}. 
In the case of the 4th order of approximation it leads to the following shadow Hamiltonian:  
\begin{equation} 
   \label{eq:shadow}
   \mathcal{\tilde H} = U + \frac{1}{2}\dot{\mathbf{x}}[M \dot{\bf{x}}] + \frac{\Delta t^2}{12}\dot{\mathbf{x}}[M \dddot{\mathbf{x}}] - \frac{\Delta t^2}{24}\ddot{\mathbf{x}}[ M \ddot{\mathbf{x}}],
\end{equation} 
where $U$ is the potential energy, $\mathbf{x}$ is the positions vector, and $M$ is the atomic mass matrix.  
The derivatives of the positions can be obtained using the finite difference approximation.
The order of approximation of modified Hamiltonians used in the simulation also affects the acceptance rates. 
Higher approximation orders provide better acceptance rates, but they also require more time to compute.

The GSHMC method consists of a series of two alternating steps: first, one integrates a short MD trajectory at constant energy, and then performs a partial momentum update.
Each of these steps can be accepted or rejected following the result of a Metropolis test where the acceptance probabilities are calculated using shadow Hamiltonians $\mathcal{\tilde H}$ instead of true Hamiltonians. 
The full algorithm can be summarized as follows:
\begin{itemize}
 \item \textit{Molecular dynamics (MD) step}: 
   \begin{itemize} 
     \item Given vectors for positions $\mathbf{x}$ and momenta $\mathbf{p}$, temperature $T$ and mass matrix $M$, integrate the Hamiltonian equations of the system:
     \begin{equation}
     \label{eq:hamiltonian_eqs}
        \begin{array}{c}
           \dot{\mathbf{p}}= -\displaystyle\frac{\partial \mathcal{H}}{\partial \mathbf{x}},
           \ \ \dot{\mathbf{x}}= \displaystyle\frac{\partial \mathcal{H}}{\partial\mathbf{p}}
         \end{array}
     \end{equation}
     with
     \begin{equation}
        \nonumber
        \label{eq:Hamiltonian}
        \begin{array}{ll}
           \mathcal{H}
           =\displaystyle\frac{1}{2}\mathbf{p}^T M^{-1} \mathbf{p} + U(\mathbf{x}),
        \end{array}
      \end{equation}
      using a symplectic method $\varPsi_{\Delta t}$ over $L$ steps with time step $\Delta t$.
      This generates a new configuration $\varPsi_{\mathcal{T}}(\bf{x},\bf{p})=(\bf{x}', \bf{p}')$, with $\mathcal{T}=L\Delta t$.

    \item Accept or reject the new configuration $(\mathbf{x}', \mathbf{p}')$ by performing a Metropolis test with the probability
       \begin{equation}
       \nonumber
           \min\left\{1,\frac
           {\exp \left( -\beta \mathcal{\tilde H}(\mathbf{x}',\mathbf{p}') \right) }
           {\exp \left( -\beta \mathcal{\tilde H}(\mathbf{x},\mathbf{p}) \right) }
           \right\},
    \end{equation}
    where $\beta=1/k_BT$ with $k_B$ being the Boltzmann constant and $\mathcal{\tilde H}(\bf{x},\bf{p})$ the shadow Hamiltonian.
 
    \begin{itemize}
       \item If accepted: save $\mathbf{x'}$ and $\mathbf{p'}$ as the current positions and momenta $(\mathbf{x},\mathbf{p})$.
       \item If rejected: restore the initial $\mathbf{x}$ and $\mathbf{p}$ and negate the momenta to ensure the stationarity of the canonical distribution.
    \end{itemize}
  \end{itemize}

 \item \textit{Partial momentum update (PMU) step}
    \begin{itemize}
       \item Generate a noise vector $\bf{u}$ from the Gaussian distribution $\mathcal N (0,\beta^{-1} M)$ as 
             \begin{equation}
             \nonumber
                  \label{Eq:Noise}
                   \mathbf{u} = \beta^{-1/2} M^{1/2} \xi, 
             \end{equation} 
             where $\xi=(\xi_1,\ldots,\xi_{3N})^T$, $\xi_i\sim \mathcal{N}(0,1)$, $i=1,\ldots,3N$ and $N$ is the system size.
      
       \item For the current positions $\mathbf{x}$  update the momenta $\mathbf{p}$ using the partial momentum update procedure:
             \begin{equation}
             \label{eq:Rotation}
                      \left(\begin{array}{c}
                      \mathbf{u'} \\
                      \mathbf{p'}
                   \end{array}\right)
                   =
                   \left(\begin{array}{cc}
                      \cos(\phi) & -\sin(\phi)\\
                      \sin(\phi) & \cos(\phi)
                   \end{array}\right)
                   \left(\begin{array}{c}
                      \mathbf{u} \\
                      \mathbf{p}
                   \end{array}\right),
              \end{equation}
              $\phi$ is a parameter taking values from $(0,\pi/2]$.

        \item Accept or reject the new momenta $\mathbf{p'}$ by performing a Metropolis test with the probability 
              \begin{equation}
              \nonumber
              \label{eq:Probability}
                    \min\left\{1,\frac
                    {\exp \left( -\beta[\mathcal{\tilde H}(\mathbf{x},\mathbf{p'})+\frac{1}{2}(\mathbf{u'})^T M^{-1} \mathbf{u'}] \right)}{
                    \exp \left( -\beta[\mathcal{\tilde H}(\mathbf{x},\mathbf{p})+\frac{1}{2}{\mathbf{u}}^T M^{-1} \mathbf{u}] \right)}
                    \right\}.
               \end{equation}
               \begin{itemize}
                     \item If accepted: save $\mathbf{p'}$ as the current momentum $\mathbf{p}$.
                     \item If rejected: restore the initial $\mathbf{p}$.
               \end{itemize}
  \end{itemize}
  
\end{itemize}
Repeat MD and PMU step for a desired number of iterations.

As the simulation is performed in a modified ensemble with respect to shadow Hamiltonians, reweighting has to be applied to calculations of statistical averages \cite{Akhmatskaya2008}. More specifically, given an observable $\Omega(\mathbf{x},\mathbf{p})$ and its values $\Omega_i$, $i=1,\ldots,K$, along a sequence of states $(\mathbf{x}_i,\mathbf{p}_i)$, $i=1,\ldots,K$, the averages $\langle \Omega \rangle$ are calculated as
\begin{equation}\label{EQ__30} 
\nonumber
\langle \Omega \rangle_K =\frac{\sum _{i=1}^{K} w_i \Omega_i}{\sum _{i=1}^{K} w_i
}  
\end{equation} 
with weight factors 
\begin{equation}\label{EQ__31} 
\nonumber
w_{i} =\exp \left[ 
  -\beta \left( \mathcal{H}(\mathbf{x}_i,\mathbf{p}_i) -
   \mathcal{\tilde H} (\mathbf{x}_i,\mathbf{p}_i) \right) 
   \right]. 
\end{equation}

\subsection{New features introduced in GSHMC for this study}

\subsubsection{Modified Adaptive Integration Approach (MAIA)}
\label{sec:MAIA}

As was pointed above, the original GSHMC method uses the leapfrog integrator and the corresponding modified Hamiltonians of arbitrary accuracy \cite{Akhmatskaya2008}. 
The leapfrog integrator is a popular choice for molecular dynamics due to the favorable combination of properties such as the second order of accuracy, the reasonably long stability limit interval, the simplicity and computational efficiency.   
Recently, Radivojevi\'{c} et al.~\cite{Radivojevic2017} demonstrated that replacing the leapfrog integrator with the one-parameter 2-stage splitting adaptive integrator specially designed for shadow Hamiltonian Monte Carlo methods may significantly improve accuracy and sampling performance of GSHMC \cite{Radivojevic2,RadivojevicThesis}. 
The authors termed this scheme the Modified Adaptive Integration Approach or MAIA.
The adaptive integrator is uniquely determined for a given simulated system and simulation time step, in such a way that the expected error in modified Hamiltonians, $\Delta \mathcal{\tilde H}$, is minimal. 
This immediately implies the best acceptance rates possible within a chosen setup since GSHMC does sample with respect to modified Hamiltonians and, therefore, the error $\Delta \mathcal{\tilde H}$ enters the Metropolis test. 

In this study, we investigate the efficiency of the MAIA method in simulations of olivine NaFePO$_4$. We briefly summarize MAIA below.

A 2-stage one-parameter splitting integrator $\psi_{\Delta t}$ of a Hamiltonian system \eqref{eq:hamiltonian_eqs} with a Hamiltonian
\begin{equation}
 \mathcal{H}(\mathbf{x},\mathbf{p})= \frac{1}{2} \mathbf{p}^T M^{-1} \mathbf{p}+U(\mathbf{x}) \equiv A+B
\label{eq:1}
\end{equation}
is defined as a composition of solution $g$-flows of partial systems $X \in \{A,B\}, \Phi_g^X$, where $g=\{b \Delta t , \Delta t/2, (1-2b) \Delta t \}$,
$\Delta t$ is a time step and $0<b<1/2$ is a parameter of the family:
\begin{eqnarray}
\nonumber
\psi_{\Delta t} &=& ( \phi^B_{b\Delta t} \circ \phi^A_{\Delta t/2} \circ \phi^B_{(1/2-b)\Delta t} ) \circ \\ 
                & &( \phi^B_{(1/2-b)\Delta t} \circ \phi^A_{\Delta t/2} \circ \phi^B_{b\Delta t} ) \equiv \Phi^1_{\Delta t/2} \circ \Phi^2_{\Delta t/2}.
\label{eq:2}
\end{eqnarray}
The maps $\Phi^1_{\Delta t/2}$ and $\Phi^2_{\Delta t/2}$ advance the solution over a first and a second halves step of length $\Delta t/2$ respectively, therefore the name 2-stage for this integrators family.

Such an integrator is symplectic as a composition of symplectic flows and reversible due to the palindromic structure of \eqref{eq:2}. A free parameter $b$ fully describes a 2-stage integrator and can be chosen accordingly to some special requirements on the properties of an integrator. 
Several 2-stage splitting integrators with the parameters $b$ fixed to some specific values are commonly used in molecular dynamics and/or Hybrid Monte Carlo methods \cite{McLachlan1995,Blanes2014}.
The most celebrated one is the Verlet/leapfrog integrator. Indeed, with $b=1/4$ both maps in \eqref{eq:2}, $\Phi^1_{\Delta t/2}$ and $\Phi^2_{\Delta t/2}$, become a velocity Verlet (VV) algorithm with a time step of $\Delta t/2$: 
\begin{multline}
\nonumber
\psi_{\Delta t} = ( \phi^B_{\Delta t/4} \circ \phi^A_{\Delta t/2} \circ \phi^B_{\Delta t/4} ) \circ 
                  ( \phi^B_{\Delta t/4} \circ \phi^A_{\Delta t/2} \circ \phi^B_{\Delta t/4} ) 
                  \\
                = \Phi^1_{\Delta t/2} \circ \Phi^2_{\Delta t/2}
                \equiv \psi^{VV}_{\Delta t/2} \circ \psi^{VV}_{\Delta t/2}.
\label{eq:3}
\end{multline}

This suggests that in order to make a fair comparison in terms of computational efficiency between an arbitrary 2-stage scheme with the parameter $b \neq 1/4$ and the Verlet integrator in its usual formulation, a 2-stage integrator \eqref{eq:2} should be run with a twice longer time step than Verlet, but for a twice shorter number of integration steps $L$, i.e. $\Delta t_\text{2-stage}=2\Delta t_\text{Verlet}$ and $L_\text{2-stage}=L_\text{Verlet}/2$. 
Some specific choices of $b$ in \eqref{eq:2} lead to the 2-stage integrators, which are capable of outperforming Verlet in accuracy and efficiency with the appropriately selected time steps as it was demonstrated in Refs. \cite{McLachlan1995,Blanes2014,Radivojevic2,RadivojevicThesis}. 
However, with an increasing time step the Verlet integrator shows better performance due to the longer stability limit interval (see for example Ref. \cite{Fernandez2016}).

The MAIA approach \cite{Radivojevic2017} provides a rational choice of an integration parameter and identifies a unique value $b^*$ of the parameter $b$ (and thus a unique integrator) for a given simulated system and a chosen time step $\Delta t$ as 
\begin{equation}
b^*=\arg \min\limits_{0<b<\frac{1}{4}} \max\limits_{0<h<\bar{h}} \rho(h,b),
 \label{eq:4}
\end{equation}
where $\rho(h,b)$ is the upper bound for the expected value of the modified energy error $\Delta = \mathcal{\tilde H}(\psi_{L h}(\mathbf{x},\mathbf{p})) - \mathcal{\tilde H}(\mathbf{x},\mathbf{p})$ with respect to the modified density $\pi(\mathbf{x},\mathbf{p})\propto e^{-\beta \mathcal{\tilde{H}}(\mathbf{x},\mathbf{p})}$, i.e. $\mathbb{E}_{\pi}(\Delta) \leq \rho(h,b)$.
Here, as before, $\mathbf{x}$ and $\mathbf{p}$ are position and momentum respectively, $\psi_{L h}$ is a 2-stage integrator advancing the numerical solution over $L$ steps, $h$ is a dimensionless time step, and $\bar{h}=\sqrt{2}\omega\Delta t$ with $\omega$ being the highest frequency of the simulated system. 
Such a choice of $b^*$ guarantees the best conservation of the modified Hamiltonians and thus the best acceptance of proposals in the GSHMC method.
Depending on the values of the highest frequency of a simulated system $\omega$, and a choice of a time step $\Delta t$, the adaptive integrator can either coincide with already known integrators with a fixed parameter, e.g. Verlet, the minimum-error integrator, ME \cite{McLachlan1995}, or BCSS \cite{Blanes2014} or be a new integrator, whose efficiency is the best under the chosen conditions.

The derivation of $\rho(h,b)$ is described in Ref.~\cite{Radivojevic2017}, whereas the formulae for modified Hamiltonians $\mathcal{ \tilde H}(\mathbf{x},\mathbf{p})$ of various orders of approximation corresponding to the multi-stage splitting integrators were obtained in Ref.~\cite{Radivojevic2,RadivojevicThesis}.
Here we only present the expressions we used in this study.

The 4th order modified Hamiltonian for 2-stage splitting integrators derived in terms of quantities available during a simulation reads as:
\begin{multline}
 \nonumber
 \mathcal{\tilde H}(\mathbf{x},\mathbf{p})= \frac{1}{2} \mathbf{p}^T M^{-1} \mathbf{p} + U(\mathbf{x}) \\
 +\Delta t^2 \left( \alpha \mathbf{p}^TM^{-1} \nabla \dot U(\mathbf{x}) + \beta \nabla U(\mathbf{x})^ T M^{-1} \nabla U(\mathbf{x}) \right),
\end{multline}
with
\begin{equation}
\nonumber
 \alpha=\frac{6b^*-1}{24},
\end{equation}
\begin{equation}
\nonumber
 \beta=\frac{6b^{*2}-6b^*+1}{12},
\end{equation}
where $\nabla \dot U(\mathbf{x})$ is the numerical time derivative of the gradient of the potential $\nabla U(\mathbf{x})$ and $b^*$ is a parameter of a system specific 2-stage integrator.
The upper bound function $\rho(h,b)$ is calculated as \cite{Radivojevic2017,Radivojevic2,RadivojevicThesis}:
\begin{equation}
\nonumber
 \rho(h,b)=\frac{(SB_h+C_h)^2}{2S(1-A_h^2)},
\end{equation}
\begin{equation}
\nonumber
  S=\frac{1+2h^2\beta}{1+2h^2\alpha},
\end{equation}
\begin{equation}
\nonumber
 A_h=\frac{h^4b(1-2b)}{4}-\frac{h^2}{2}+1,
\end{equation}
\begin{equation}
\nonumber
 B_h=-\frac{h^3(1-2b)}{4}+h,
\end{equation}
\begin{equation}
\nonumber
 C_h=-\frac{h^5b^2(1-2b)}{4}+h^3b(1-b)-h.
\end{equation}
 
Importantly, finding the appropriate parameter $b^*$ in \eqref{eq:4} can be done at the pre-processing stage of the simulation.
Therefore, the procedure does not introduce any computational overhead. 
Additionally, the method is available for constrained and unconstrained dynamics and it is thus applicable to a broad range of problems.  

In Section \ref{sec:numerical} we compare performance of GSHMC achieved using two different integration schemes, the velocity Verlet and MAIA, for a range of time steps and lengths of MD trajectories. We find that using the MAIA integrators may improve performance of the original GSHMC method by a factor as high as 2.   

\subsubsection{Randomized Shell Mass Generalized Shadow Hybrid Monte Carlo (RSM-GSHMC)}

One important drawback of the adiabatic dynamics core-shell approach is its potential negative effect on the simulated kinetic properties due to the introduction of a shell mass. 
Previous studies demonstrated that with the careful choice of a shell mass and a time step such an effect becomes negligible \cite{Mitchell1993,Lindan1993}. There is not, however, a clear criterion for choosing these parameters, and finding the appropriate parameter values is a matter of trial and error. 

In this paper, we propose to take an advantage of the flexibility of the GSHMC method in order to smooth the undesired effect of the shell mass on the kinetics of a simulated system. 
The flexibility we refer to in this context is the possibility to vary on the fly in a rigorous manner the simulation parameters in GSHMC \cite{Radivojevic2,RadivojevicThesis}. 
This can be done before starting each new molecular dynamics trajectory, i.e. on each Monte Carlo step, by randomizing the simulation parameters around pre-assigned fixed values.  These parameters can be selected independently from a chosen distribution. The randomization helps to avoid some bad combinations of fixed values that might lead to accuracy or performance degradation such as slow convergence and non-ergodicity. 
Based on this idea we introduced randomization of a shell mass in the GSHMC method.

We implemented the mass randomization as a part of the momentum update step. 
Before updating the momenta, we redistribute a fraction of the atomic mass between core and shell, keeping the total mass constant:
\begin{equation}
\begin{array}{c}
 m_{c,i} = m_{c,0} - \lambda_i r
 \\
 m_{s,i} = m_{s,0} + \lambda_i r
\end{array}
\label{eq:randomize}
\end{equation}
where $m_{c,i}$ and $m_{s,i}$ are the core and shell masses at Monte Carlo step $i$, $m_{c,0}$ and $m_{s,0}$ are their respective initial values, $r$ is the amount of mass that we want to randomize, and $\lambda_i$ is a random number generated from a uniform distribution $\mathcal{U}(0,1)$ at step $i$.

It is important to notice that $m_{s,0}$ has to be large enough to ensure the stability of the numerical integrator (its minimum value will depend on the time step used in the simulation). 
For a discussion about how to choose $m_{s,0}$ see Ref. \cite{Mitchell1993}. On the other hand, $r$ should not be bigger than $m_{c,0}/2$, as that could lead to a situation in which the shell is actually heavier than the core. 
Having these constraints is enough to rigorously implement the algorithm. 
However, the optimal choice of $r$ remains empirical.
Below we summarize the modified momentum update step in RSM-GSHMC. 

\begin{itemize}
       \item Given the mass matrix $M$, generate a randomized mass matrix $M'$ by applying the randomization described in \eqref{eq:randomize} to the core and shell particles.
       \item Generate a noise vector $\mathbf{u}$ from the Gaussian distribution as in the original GSHMC:
             \begin{equation}
             \nonumber
                   \mathbf{u'}=\beta^{-1/2} M'^{1/2}\xi.
             \end{equation}
       \item Adjust the current momenta $\mathbf{p}$ to the new masses:
             \begin{equation}
             \nonumber
                  \mathbf{p'}=M'^{1/2} M^{-1/2}\mathbf{p}. \\
             \end{equation}
      \item Update the candidate momenta $\mathbf{p'}$ using the partial momentum update procedure:
             \begin{equation}
             \nonumber
                      \left(\begin{array}{c}
                      \mathbf{u''} \\
                      \mathbf{p''}
                   \end{array}\right)
                   =
                   \left(\begin{array}{cc}
                      \cos(\phi) & -\sin(\phi)\\
                      \sin(\phi) & \cos(\phi)
                   \end{array}\right)
                   \left(\begin{array}{c}
                      \mathbf{u'} \\
                      \mathbf{p'}
                   \end{array}\right).
              \end{equation}
        \item Accept or reject the new momenta $\mathbf{p''}$ by performing a Metropolis test with the probability
              \begin{equation}
              \nonumber
                    \min\left\{1,
                    \frac
                    {\exp \left(-\beta[\mathcal{\tilde H}(\mathbf{x}, \mathbf{p''} )+\frac{1}{2}(\mathbf{u''})^T M'^{-1} \mathbf{u''} ] \right) }
                    {\exp \left(-\beta[\mathcal{\tilde H}(\mathbf{x}, \mathbf{p'})+\frac{1}{2} (\mathbf{u'})^T M'^{-1} \mathbf{u'} ] \right) }
                    \right\}.
               \end{equation}
               \begin{itemize}
                     \item If accepted: save $\mathbf{p''}$ as the current momenta $\mathbf{p}$.
                     \item If rejected: restore the initial $\mathbf{p}$.
               \end{itemize}
  \end{itemize}

\subsubsection{Implementation - MultiHMC}

The GSHMC method was implemented in the open source molecular dynamics software GROMACS \cite{Hess2008}, version 4.5.4. 
Details of this implementation can be found in Refs. \cite{Escribano2013} and \cite{Fernandez2014}. 
GROMACS was chosen for its popularity, computational efficiency, and effective parallelization. 
The implementation of the GSHMC method was done in a self-contained manner, respecting the parallel scalability and introducing almost no computational overhead.
The same software package was used for the implementation of 2-stage integrators \cite{Fernandez2016} and, in particular, for the MAIA method \cite{Radivojevic2017}. We call the resulting package \textsf{MultiHMC-GROMACS} and it is available for public use under the GNU Lesser General Public License.
All the classical atomistic simulations in this work were performed with the \textsf{MultiHMC-GROMACS} software.

Additionally, the randomized mass algorithm for the adiabatic core-shell model was implemented in the same software package as a single function call inside the partial momentum update step.

\section{Numerical Experiments}
\label{sec:numerical}

In this section we present a series of numerical experiments performed to validate our computational model and to evaluate the performance of the proposed sampling approach. 
To this end we used four different atomistic simulation methods: MD (CS-min), MD (CS-adi), GSHMC and RSM-GSHMC. 
In addition, we compared the results with available experimental data and assessed the accuracy of the underlaying force field by performing some groundstate DFT calculations.

\subsection{DFT calculations}

The total energies were computed using the projected augmented wave (PAW) method \cite{Blochl1994,Kresse1999} within the PBE generalized gradient approximation (GGA) \cite{Perdew1996} as implemented in the VASP package version 5.3.3 \cite{Kresse1996b}.
The GGA+U approach, in which an effective  Hubbard U-like term  is added to exchange-correlation functional, was required to correctly account for the electronic correlation of iron $3d$ electrons \cite{Zhou2004a}.
We used a $U$ value of 4.3~eV as suggested for NaFePO$_4$ in other works \cite{Saracibar2016,Boucher2014,Lu2013}.
An energy cutoff of 600 eV and a proper $k$-point mesh were used to ensure that the total energies had converged within 5~meV per formula unit (f.u.). The geometry optimization was considered converged when forces on the atoms for each component became smaller than 0.02~eV/\AA.

The surface energies of different possible terminations for NaFePO$_4$ were computed following the approach outlined by Wang et al.~\cite{Wang2007} for its lithium counterpart. 
In Table \ref{tab:surfener} the DFT results obtained for surface energies are shown in comparison with the ones reported using the interatomic potential chosen for this study \cite{Whiteside2014}. 
The tested methods provide similar values of the surface energies and close trends in the surfaces stability order. The agreement is very good considering the fact that the classical model was obtained by fitting to bulk structural properties only. These results support our choice of the interatomic potential for atomistic simulations carried out in this study.

\begin{table}
\centering
	\caption{NaFePO$_4$ surface energies ($\gamma$) for different terminations considered after relaxation, as determined by DFT calculations in the present work ($\gamma_\text{DFT}$) and with classical interatomic potentials ($\gamma_\text{FF}$) \cite{Whiteside2014}.} 
		\begin{tabular}{ c c c } \toprule
		Surface      & $\gamma_\text{DFT}$ (J/m$^2$)     & $\gamma_\text{FF}$ (J/m$^2$) \\ \midrule
		(010)        & 0.51              & 0.52                  \\ 
		(201)        & 0.59              & 0.63                  \\
		(101)        & 0.60              & 0.74                  \\
		(100)        & 0.67              & 0.68                  \\  
		(110)        & 0.70              & 0.54                  \\
		(111)        & 0.97              & 0.68                  \\  
		(001)        & 1.15              & 0.90                  \\
		\bottomrule
	\end{tabular}
    \label{tab:surfener}
\end{table}

\subsection{MD simulations}

We considered two different systems: a fully sodiated NaFePO$_4$ and a partially sodiated Na$_{2/3}$FePO$_4$. 
The latter was chosen as the most stable compound reported by Saracibar et al.~\cite{Saracibar2016} for that composition, which corresponds to a stable ordered superstructure in the Na$_x$FePO$_4$ ($0\leq x\leq 1$) phase diagram \cite{Galceran2014,Boucher2014,CasasCabanas2012}.
The unit cell of Na$_{2/3}$FePO$_4$ contains 12 f.u., i.e. 80 atoms.

For bulk NaFePO$_4$ we built a model system based on a ($6 \times6 \times6$) supercell containing 864 NaFePO$_4$ f.u. (10368 particles in total including the Fe and O shells). For Na$_{2/3}$FePO$_4$ we used a ($6 \times 3 \times 2$) supercell with 432 Na$_{2/3}$FePO$_4$ f.u. (5040 particles).
In both cases the force field parameters were those presented in Section \ref{sec:interatomic}, with a cutoff of 12 \AA\ for electrostatics and periodic boundary conditions applied in the three dimensions. For the partially sodiated case, the extra charge in the system due to removing 1/3 of Na atoms was compensated by averaging the net charge on Fe atoms as was previously suggested in the similar study for LiFePO$_4$ \cite{Tealdi2012}.

All of the simulations were initially equilibrated with a 50~ps run using an NPT ensemble with a specified target temperature ($T$) and pressure $P=1$~bar. We employed the Berendsen thermostat \cite{Berendsen1984} and the Andersen barostat \cite{Andersen1980}.

Na-ion diffusion events require the presence of vacant Na sites, thus only the partially sodiated system was used for the computation of diffusion coefficients.
The production runs in these cases were performed in an NVT ensemble at temperatures between 10~K and 700~K using the Berendsen thermostat. 

The velocity Verlet integrator was used for all MD simulations. 
The optimal choice of the time step in MD is discussed in Section \ref{subsec:acc_and_perf}.  

\subsection{GSHMC simulations}

We tested two versions of the GSHMC method: the original approach and the RSM-GSHMC.
We used the same MD setup as described above in the MD runs of the GSHMC methods with two exceptions. 
The MD trajectories were run in an NVE ensemble, thus no thermostats were involved, and the MAIA integrator was chosen instead of velocity Verlet for most of the tests. 

The velocity Verlet integrator was coupled  with the original GSHMC method in the parameters refining procedure in order to compare its performance with respect to the MAIA integrators.
As in the case of MD simulations, the choice of the time step will be discussed in the following sections. The parameter $\phi$ in Eq. \eqref{eq:Rotation} was fixed to 0.2. 
The number of integration steps was 500 for MAIA and 1000 for velocity Verlet.
The fourth order modified Hamiltonian was used in all tests.

\subsection{Validation}
\label{subsec:validation}

First, we verified that the underlying force field used in this study, provides reliable results when employed for dynamical simulations. 
In addition, we wanted to check that the proposed simulation techniques with the chosen simulations settings are capable to accurately reproduce the properties of olivine NaFePO$_4$. 

To this end, we calculated the lattice constants of the fully sodiated NaFePO$_4$ at $T=300$~K and $P=1$~bar, based on production runs of 0.5~ns using GSHMC, RSM-GSHMC, MD (CS-min) and MD (CS-adi).
The results are shown in comparison with the experimental data \cite{Moreau2010} and the DFT-based calculations in Table \ref{tab:lattice_const}.
We found that all the methods yield very similar lattice constants, with relative differences less than 2\%.

\begin{table*}
\centering
\caption{Computed lattice constants of olivine NaFePO$_4$ at 300~K using different approaches. Experimental values also shown are taken from Ref.~\cite{Moreau2010}.}
\label{tab:lattice_const}      
\begin{tabular}{lllllll}
\toprule
Parameter (\AA) & Exp.  & DFT   & MD (CS-min) & MD (CS-adi) & GSHMC & RSM-GSHMC  \\
\midrule
a               & 10.41 & 10.52 & 10.34      & 10.38          & 10.39 & 10.40 \\
b               &  6.22 &  6.27 &  6.16      &  6.19          &  6.19 &  6.19 \\
c               &  4.95 &  4.99 &  4.90      &  4.92          &  4.92 &  4.91 \\
\bottomrule
\end{tabular}
\end{table*}

\begin{figure}
  \includegraphics[width=0.5\textwidth]{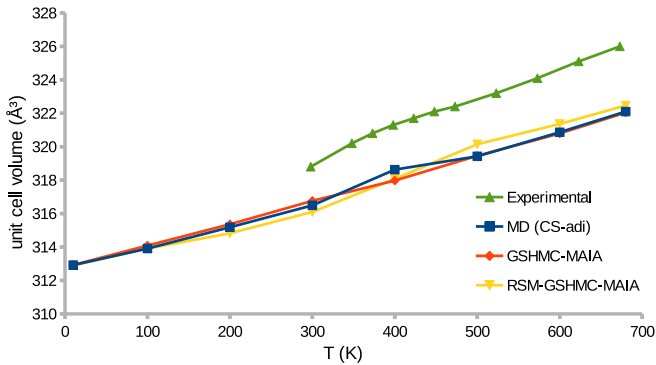}
\caption{Thermal expansion for olivine NaFePO$_4$ calculated using MD, GSHMC, RSM-GSHMC in an NPT ensemble with the Andersen barostat. Experimental values are taken from Ref.~\cite{Moreau2010}.
}
\label{fig:thermal_exp}       
\end{figure}

We also considered the thermal expansion of bulk NaFePO$_4$ to evaluate the suitability of the force field. 
We computed the volume expansion of the unit cell as a function of temperature by performing simulations under an NPT ensemble.
The pressure was maintained using the Andersen barostat.
The target temperature was controlled by using the Berendsen thermostat in MD, while GSHMC keeps $T$ constant by design. 
We considered temperatures between 10~K and 700~K, making sure that the box size and the potential energy were completely stabilized before measuring the volume. 
The resulting thermal expansion for NaFePO$_4$ is shown in Figure \ref{fig:thermal_exp} along with the experimental results of Moreau et al.~\cite{Moreau2010}. 
The three tested methods yield similar slopes and the small difference observed with respect to the experimental values is negligible (relative variations are less than 1\%).
Therefore, we can conclude that the model combined with the simulation methods under study properly accounts for the thermal expansion of olivine NaFePO$_4$.

As a final validation test we present in Table \ref{tab:valid_aver} the average values for potential ($U$) and kinetic energies ($K$), temperatures, and two structural parameters, the angles between the bonded O-P-O species ($\theta_\mathrm{O-P-O}$) and their corresponding P-O distances ($d_\mathrm{P-O}$).
As can be seen from Table \ref{tab:valid_aver}, the randomized mass algorithm, RSM-GSHMC, provides the best agreement with the experimental data, which may imply the positive effect of the randomization of a shell mass on the overall accuracy of the core-shell adiabatic model.

\begin{table*}
\centering
\caption{Average values for temperature ($T$), potential energy ($U$), kinetic energy ($K$), O-P-O angles ($\theta_\mathrm{O-P-O}$) and P-O internuclear distances ($d_\text{P-O}$). Experimental values also shown are taken from Ref.~\cite{Moreau2010}.
}
\label{tab:valid_aver}       
\begin{tabular}{llllll}
\toprule
Method     & T (K)  & U (kJ/mol) & K (kJ/mol) & $\theta_\text{O-P-O}$ (deg.) & $d_\text{P-O}$ (nm) \\
\midrule
MD (CS-min) & 297.02 & -1.01x10$^7$ & 1.09x10$^4$ & 108.29      & 0.150 \\
MD (CS-adi) & 299.98 & -1.03x10$^7$ & 1.63x10$^4$ & 108.26      & 0.151 \\
GSHMC      & 296.40 & -1.03x10$^7$ & 1.60x10$^4$ & 108.30      & 0.149 \\
RSM-GSHMC  & 298.64 & -1.03x10$^7$ & 1.61x10$^4$ & 108.32      & 0.156 \\
Experiment & 300.00 &           -- &         --  & 109.47      & 0.155 \\
\bottomrule
\end{tabular}
\end{table*}

\begin{figure}
  \includegraphics[width=0.5\textwidth]{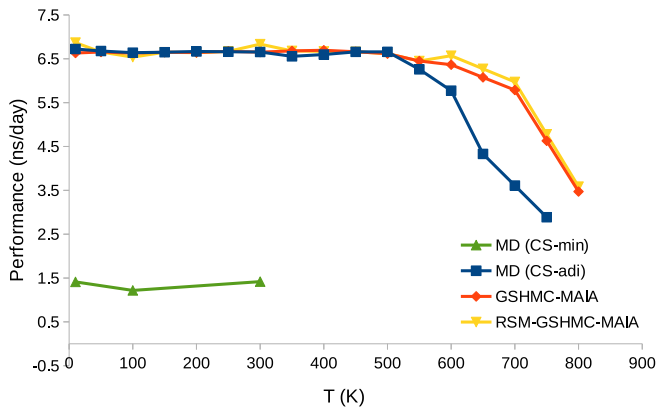}
\caption{Computational performance for all the considered methods at different temperatures.}
\label{fig:performance_ns_day}
\end{figure}

In Figure \ref{fig:performance_ns_day} we show the computational performance measured in nanoseconds per day for the four methods considered. 
All simulations were run in parallel on 8 cores on the same computational server.
In terms of performance, the adiabatic core-shell approach offers a great advantage that outweighs any marginal loss of accuracy. 
The significantly lower performance observed with the MD (CS-min) scheme is due to the big overhead introduced by the search of optimal shell positions at each time step.
The loss of performance registered at temperatures over 500~K is a consequence of using a smaller time step, which was found necessary to keep the simulations stable at such high temperatures.
The GSHMC approaches always achieved a higher performance because their increased numerical stability allowed the use of longer time steps.
For temperatures below 500~K the time steps were set to 1.15~fs for MD (CS-adi) and 2.3 fs for GSHMC methods whereas for higher temperatures they had to be reduced to 0.5~fs and 2~fs, respectively.

\subsection{Accuracy and sampling performance}
\label{subsec:acc_and_perf}

In order to include in our tests the calculation of Na self-diffusion coefficients we chose the partially sodiated Na$_{2/3}$FePO$_4$ as a benchmark system.   
The diffusion coefficients are notoriously difficult to determine from dynamical simulations because they require considerably long runs to reach convergence. 
In this work they were derived from the mean square displacement of Na-ions using the Einstein relation 
\begin{equation}
\langle|\mathbf{x}_\textrm{Na}(t+\tau)-\mathbf{x}_\textrm{Na}(t)|^2\rangle = 6D \tau,
\end{equation}
where the term on the left side, which is the squared displacement of a Na-ion during an integrated interval $\tau$, is proportional to the Na self-diffusion (or diffusion) coefficient ($D$) and $\tau$.

\begin{figure*}
	\includegraphics[width=.5\textwidth]{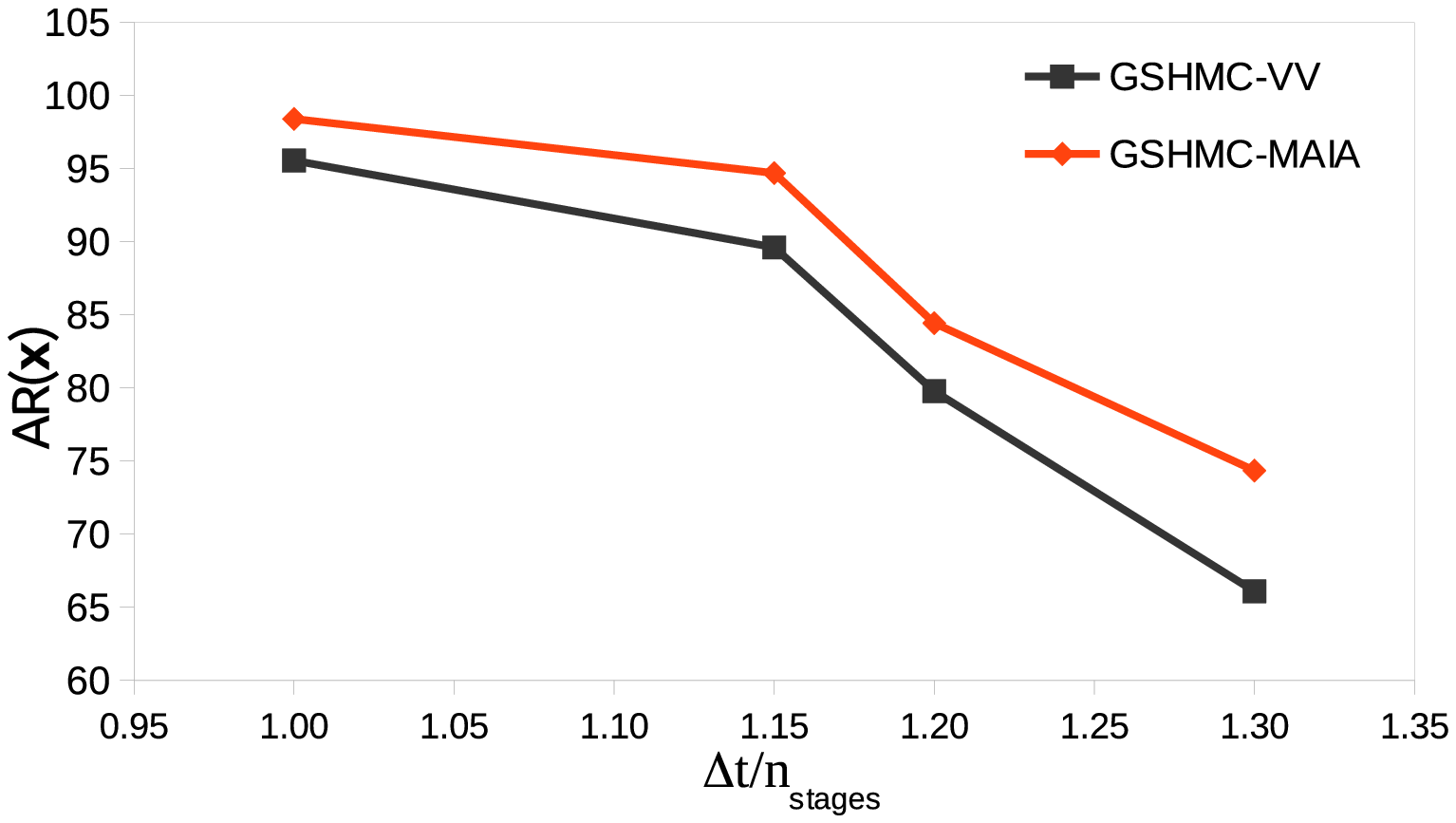}\includegraphics[width=.5\textwidth]{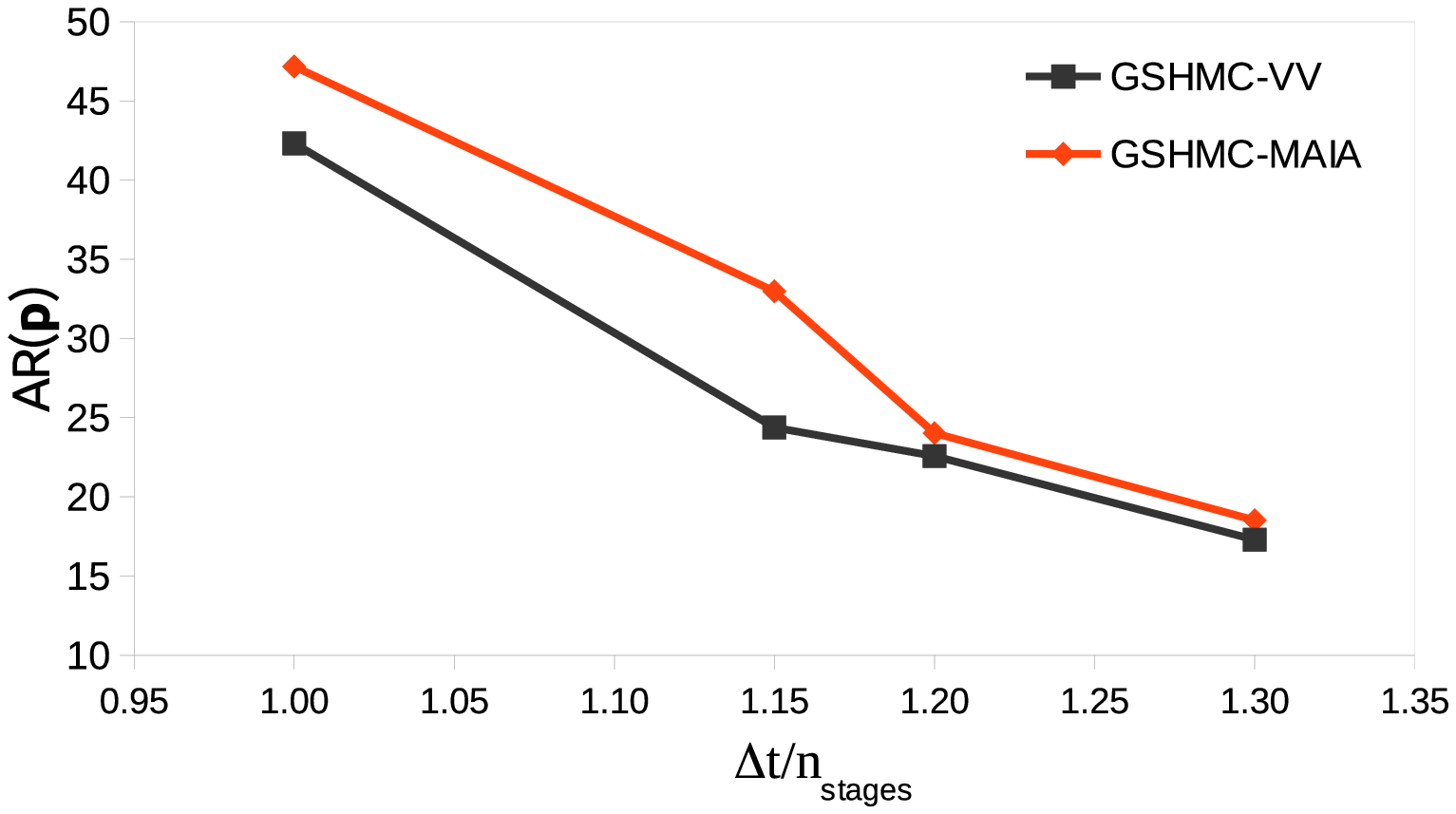}
	\caption{Acceptance rates for positions (left) and momenta (right) for GSHMC simulations using MAIA and velocity Verlet integrators at $T=300$~K.}
	\label{fig:acceptance_rates}      
\end{figure*}

In what follows, all reported properties are results of averaging over five different production runs of 2~ns each, unless stated otherwise.   

The first step for optimizing the accuracy and performance of the novel approaches in prediction of various properties of the system of interest, is to find the best combination of numerical integrators and simulation parameters to be used. We begin with measuring the sampling efficiency of GSHMC in two different scenarios, namely when the method is combined  with the new MAIA integrator or when it uses the standard velocity Verlet.

For these experiments we chose a time step in the MAIA integrator twice longer than the one in the velocity Verlet case presented in its usual, 1-stage formulation. 
However, since MAIA performs two force evaluations per each integration step in contrast to only one in velocity Verlet, a number of integration steps in MAIA has been chosen to be half as many as in velocity Verlet. 
Such a setup equalizes the computational effort required for each integrator and makes the comparison fair (see Section \ref{sec:MAIA} for the detailed explanation). 
For simplicity, we use from now on an effective time step, defined as $ \Delta t/n_\mathrm{stages}$, where $n_\mathrm{stages}$ is equal to either 1 for velocity Verlet or to 2 for MAIA.

One can evaluate the influence of the tested integrators on the sampling performance by looking at the acceptance rates (AR) for the positions and momenta Metropolis tests. 
Ideally, the AR for positions in GSHMC should be close to 100\%, to minimize the time spent on computing trajectories that are finally rejected.
In Figure \ref{fig:acceptance_rates} we show the AR for positions and momenta when using the two integrators for a range of integration time steps ($\Delta t$). 
As follows from Figure \ref{fig:acceptance_rates}, MAIA always leads to better acceptance rates, as for positions as for momenta, than can be achieved with velocity Verlet.

\begin{figure*}
  \includegraphics[width=.5\textwidth]{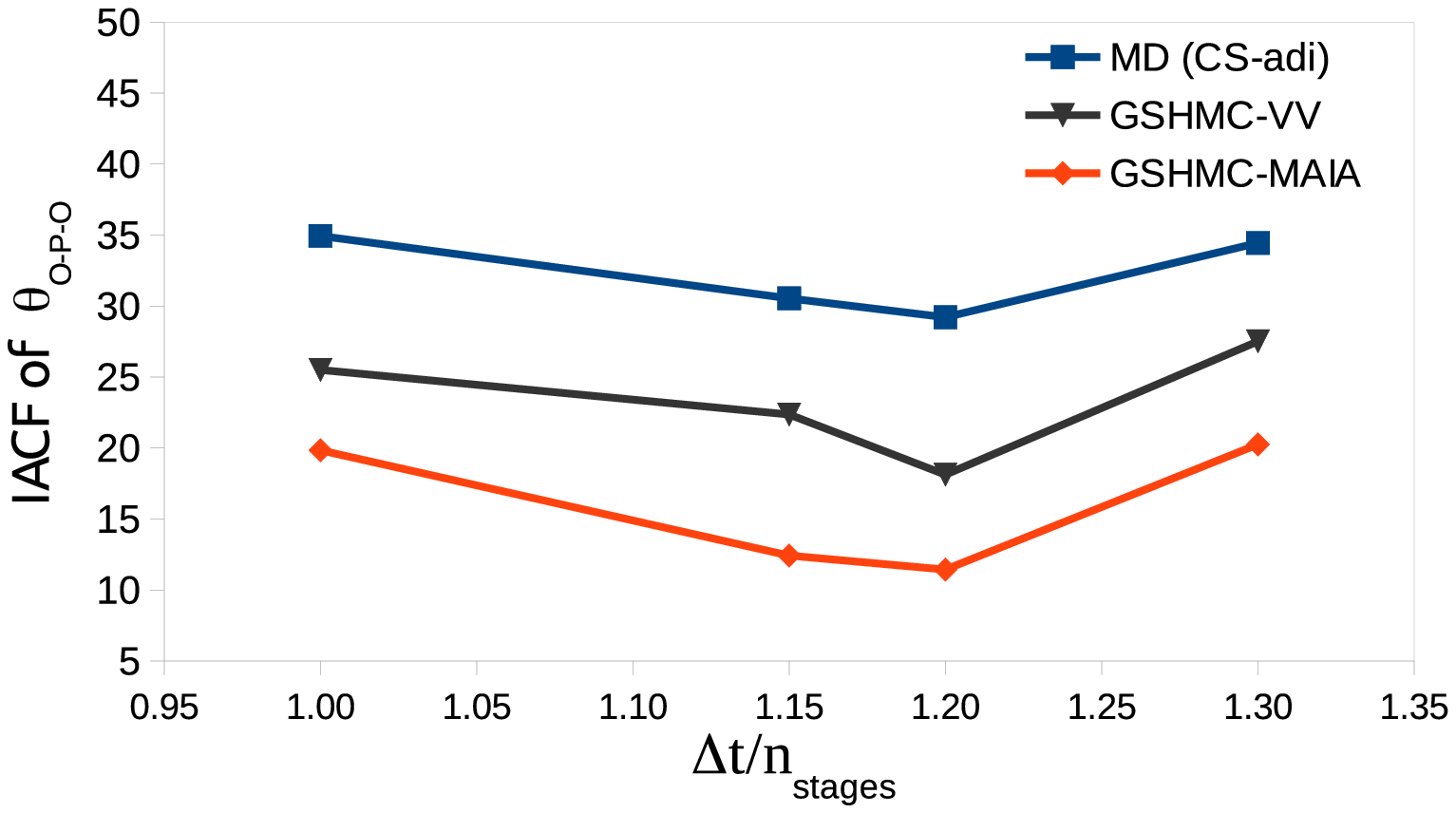}\includegraphics[width=.5\textwidth]{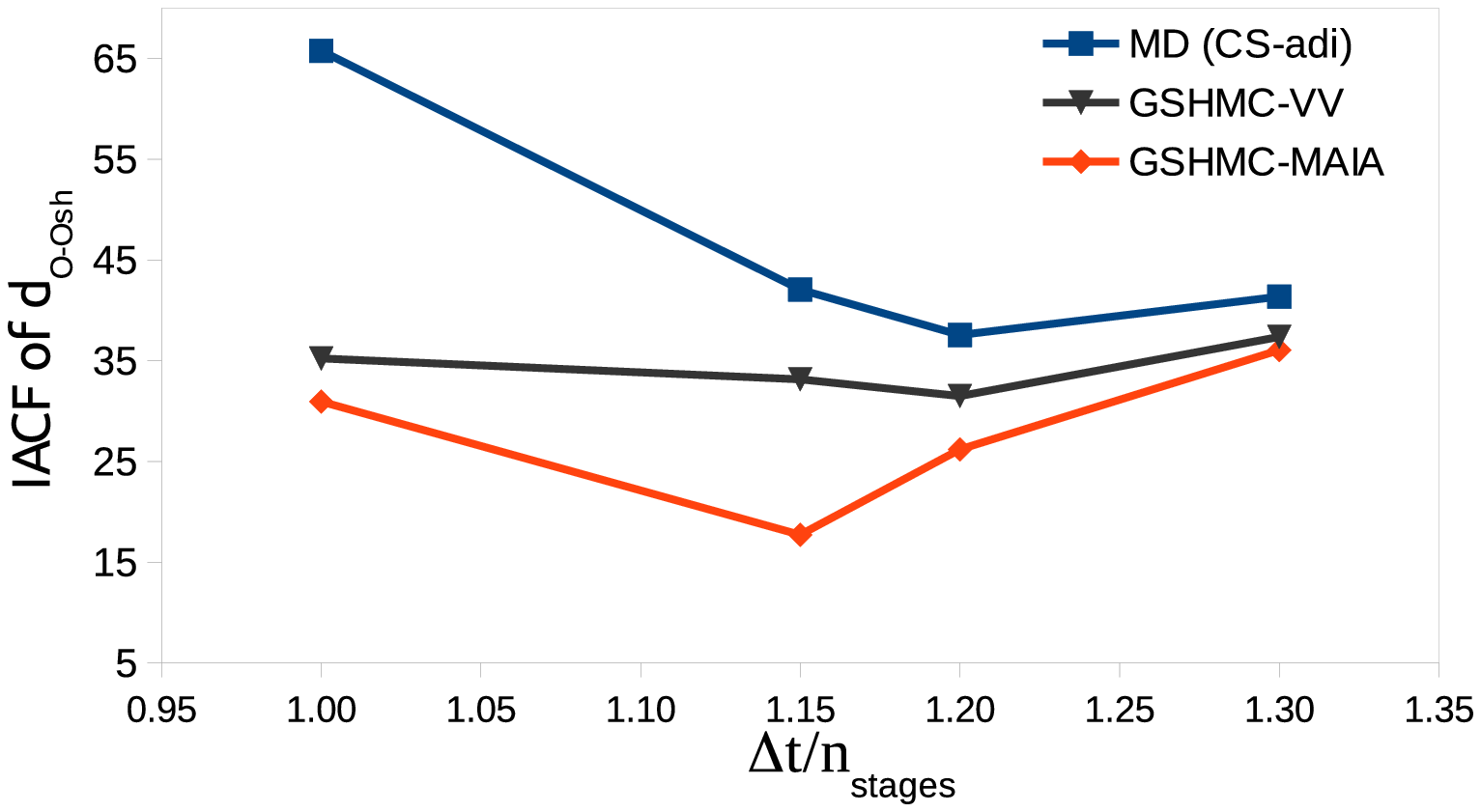}
  \includegraphics[width=.5\textwidth]{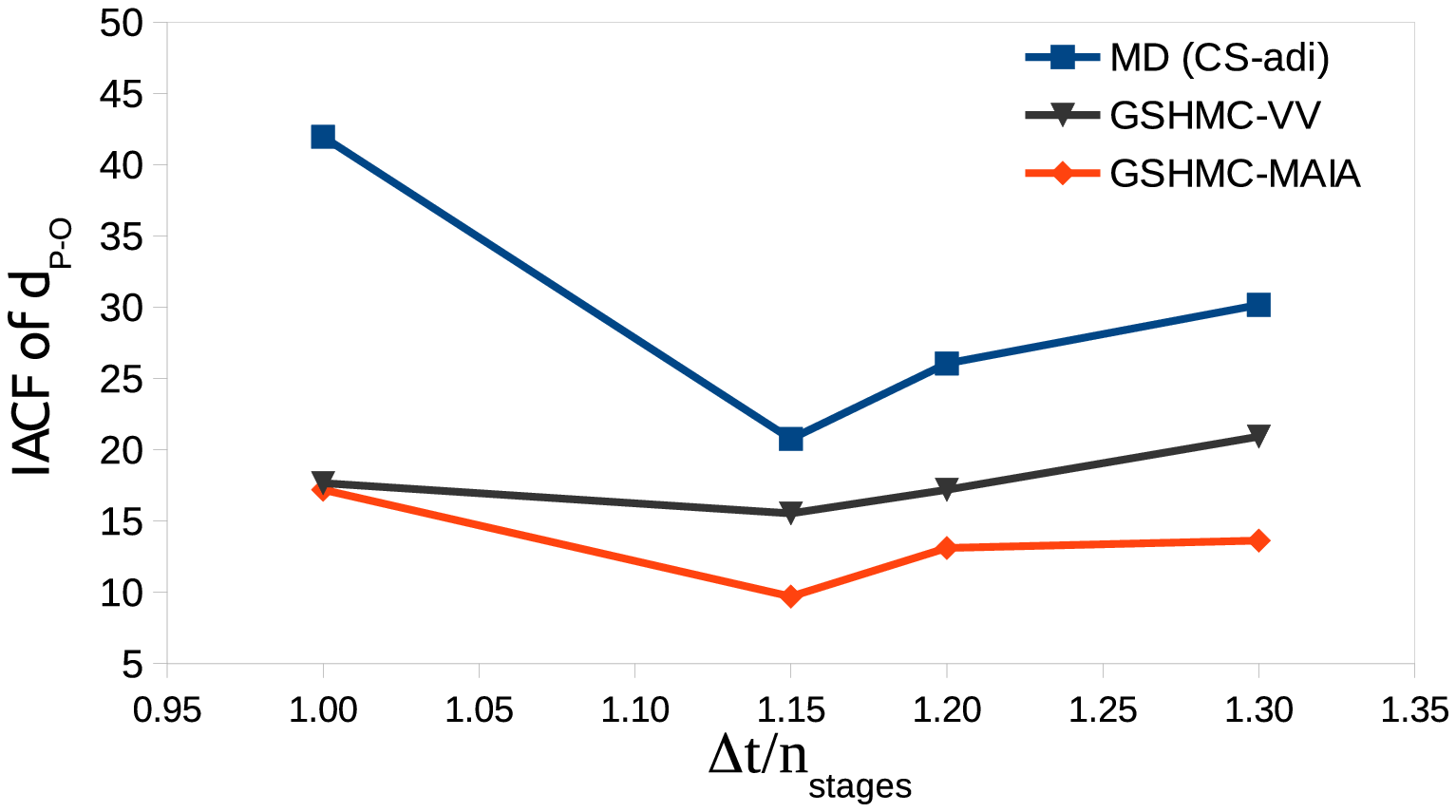}\includegraphics[width=.5\textwidth]{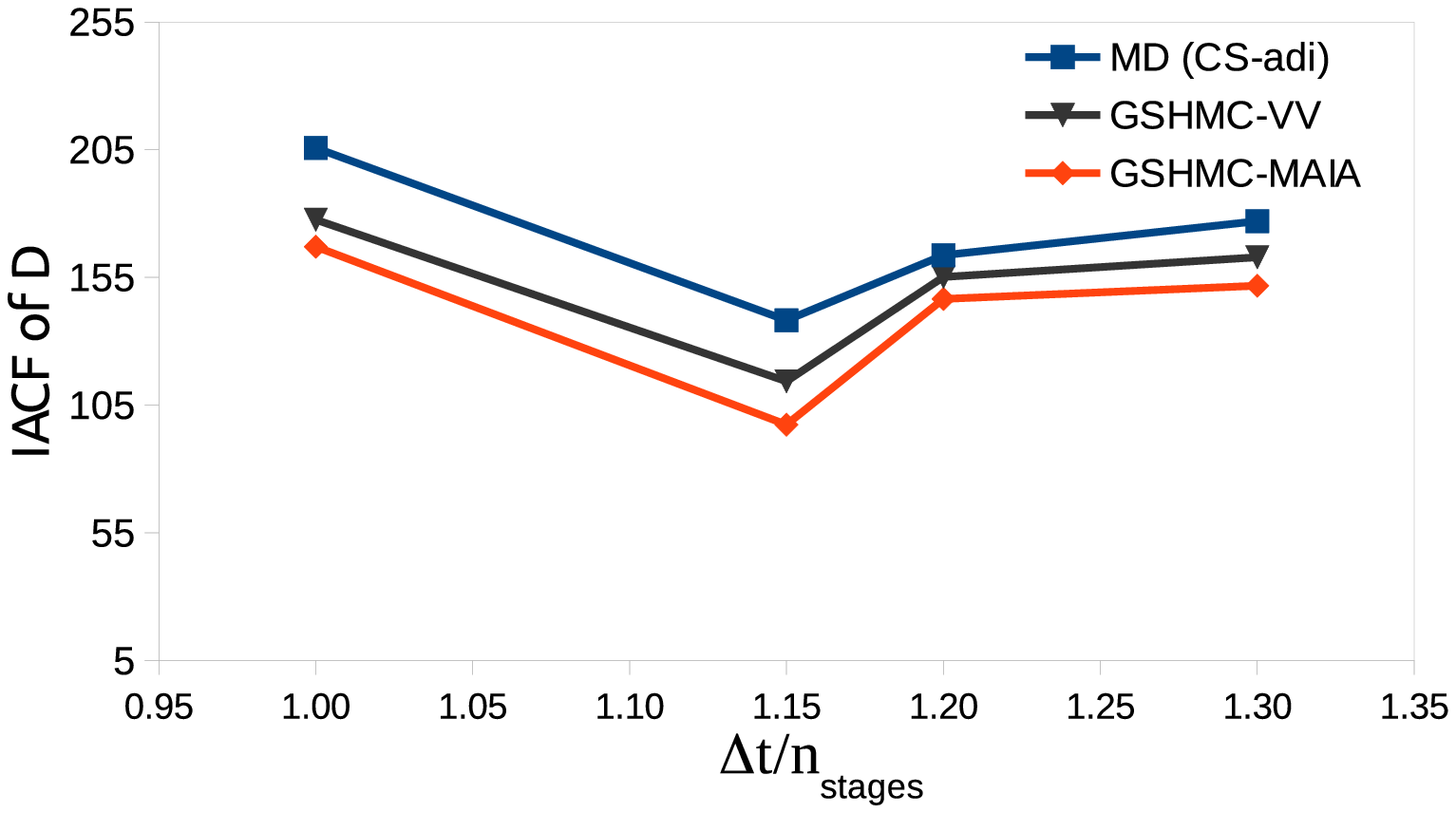}  
\caption{Integrated autocorrelation functions for the diffusion coefficient and structural parameters at $T=300$~K.}
\label{fig:IACFs}      
\end{figure*}

Another way to compare the sampling efficiency is to calculate integrated autocorrelation functions (IACF), defined as:
\begin{equation}
 \mathrm{IACF}=\sum\limits_{l=0}^{K'} ACF(\tau_l),
\end{equation}
where $ACF(\tau_l)$, $l=0,...,K'<K$ is the standard autocorrelation function for the time series $\Omega_k$ of $K$ samples, $k=1,...,K$, with the normalization
\begin{equation}
\nonumber
 ACF(\tau_0)=ACF(0)=1. 
\end{equation}
The IACF gives a quantitative measure of time required, on average, for generating a non-correlated sample, and thus lower values of IACF imply more efficient sampling. 

In Figure \ref{fig:IACFs} we present the IACF values for several properties of the system obtained with the MD and GSHMC methods for different effective time steps. 
The latter was combined with velocity Verlet (GSHMC-VV) and MAIA (GSHMC-MAIA). 
Clearly, the combination of GSHMC with MAIA always produces the lowest IACF values, which translates into a more efficient sampling. 
On the other hand, plotting IACF as a function of the effective time step, helps to reveal the influence of the time step on the overall performance and suggest a way to choose the optimal one. 
More specifically, we found that the best performance was observed for all the simulation methods at the effective time step of 1.15~fs and thus the rest of the tests were performed with this value. 
Also, since the GSHMC-MAIA combination provided the best sampling efficiency we proposed this setup for future studies.

\begin{figure}
  \centering
  \includegraphics[width=.5\textwidth]{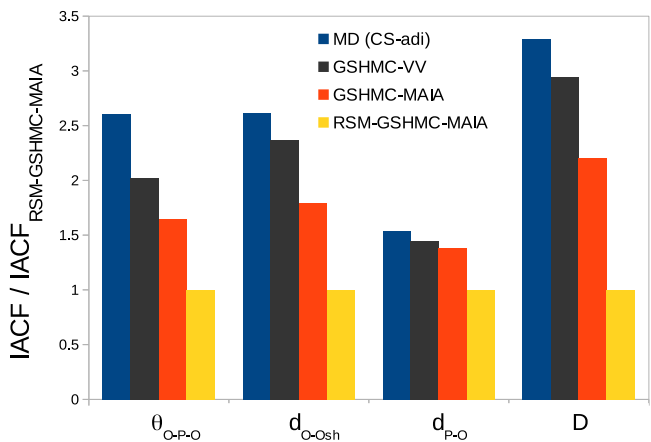} 
\caption{Relative IACF with respect to RSM-GSHMC-MAIA for structural properties and diffusion coefficients obtained with the optimal simulation parameters at $T=300$~K.}
\label{fig:IACF_bars}       
\end{figure}

Once we chose the proper settings for the MD and GSHMC methods, longer 4~ns simulations at constant volume and temperature (NVT) were performed. 
In Figure \ref{fig:IACF_bars} we plot the relative IACF for the structural parameters and self-diffusion observed with MD (CS-adi), GSHMC-VV and GSHMC-MAIA at 300K with respect to the corresponding IACF values obtained with RSM-GSHMC-MAIA.
Clearly, the best performance is obtained with the RSM-GSHMC-MAIA simulations for all simulated properties (lowest IACF values, up to 3.3 times better than in MD).
This is a very promising result, especially for computing self-diffusion coefficients in solid bulk materials.

\begin{figure}
  \centering
  \includegraphics[width=.5\textwidth]{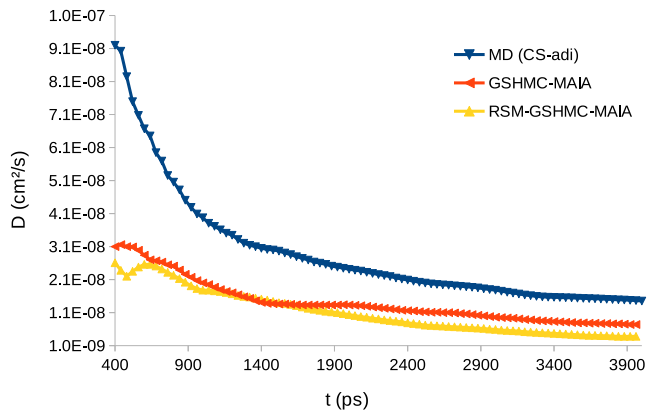} 
\caption{Diffusion coefficient convergence at $T=300$~K for MD, GSHMC and RSM-GSHMC methods.}
\label{fig:diff_conv}      
\end{figure}

The enhanced sampling of the Na-ion self-diffusion observed with RSM-GSHMC in Figure \ref{fig:IACF_bars} implies shorter integration times required for obtaining the converged self-diffusion value. 
Figure \ref{fig:diff_conv} monitors the average self-diffusion obtained with MD, GSHMC and RSM-GSHMC at $T=300$~K with increasing simulation time up to 4~ns. 
Though convergence is not fully achieved with any of the methods, the GSHMC-based, and especially RSM-GSHMC, demonstrate clear signs of convergence after 3~ns of simulation.

Next, we investigated the performance of MD, GSHMC and RSM-GSHMC in the range of temperatures by running a series of NVT simulations at temperatures between 10~K and 700~K.  As before (see section \ref{subsec:validation}), the integration time steps had to be reduced for temperatures greater than 500~K for all methods. 

We introduced a variable $X$ that measures the sampling performance by taking into account both the effective time step $\Delta t/n_\mathrm{stages}$ and the IACF as:
\begin{equation}
X=\frac{\Delta t/n_\mathrm{stages}}{\mathrm{IACF}}.
\end{equation}

\begin{figure}
 \includegraphics[width=.5\textwidth]{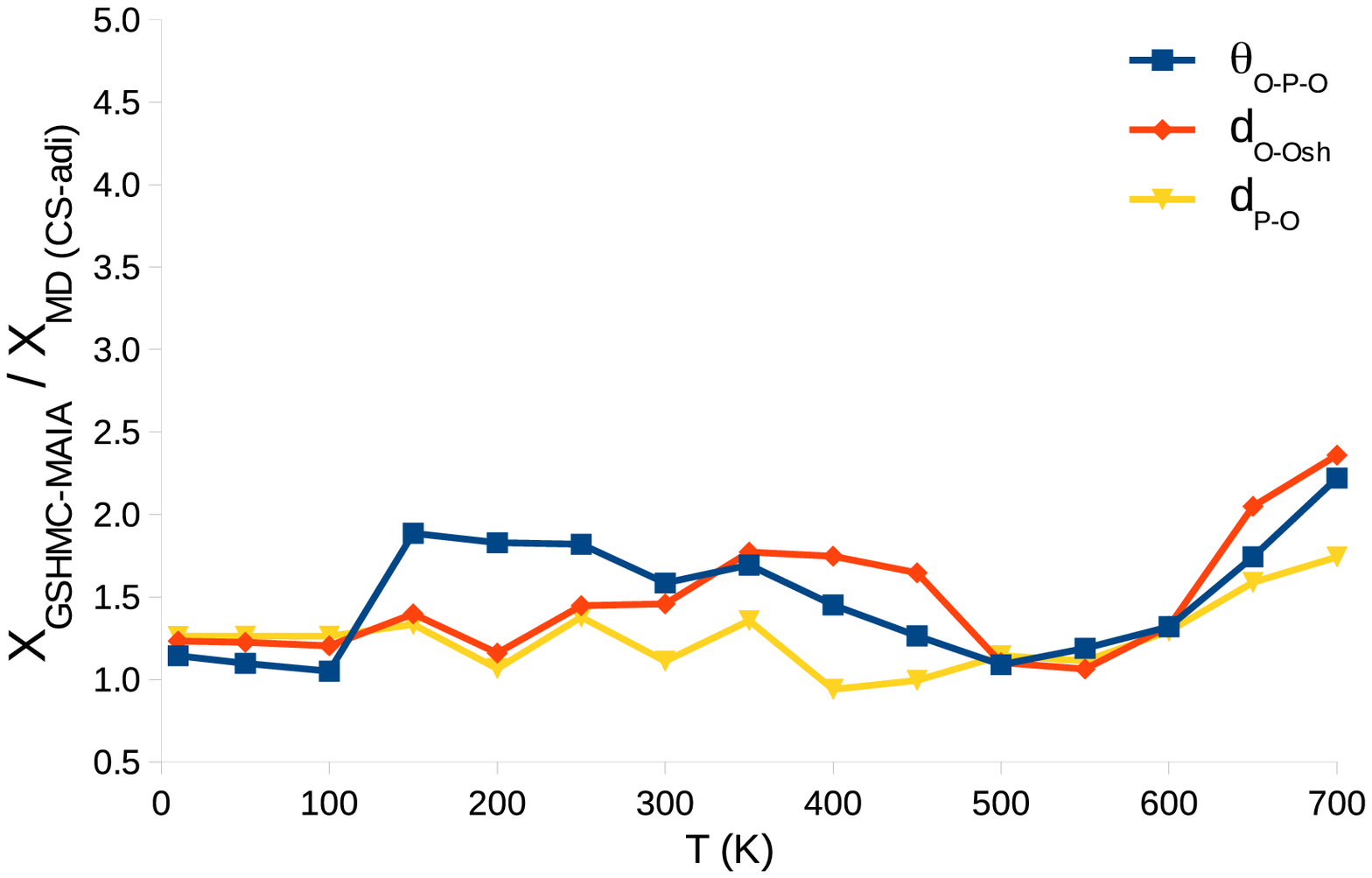}
 \includegraphics[width=.5\textwidth]{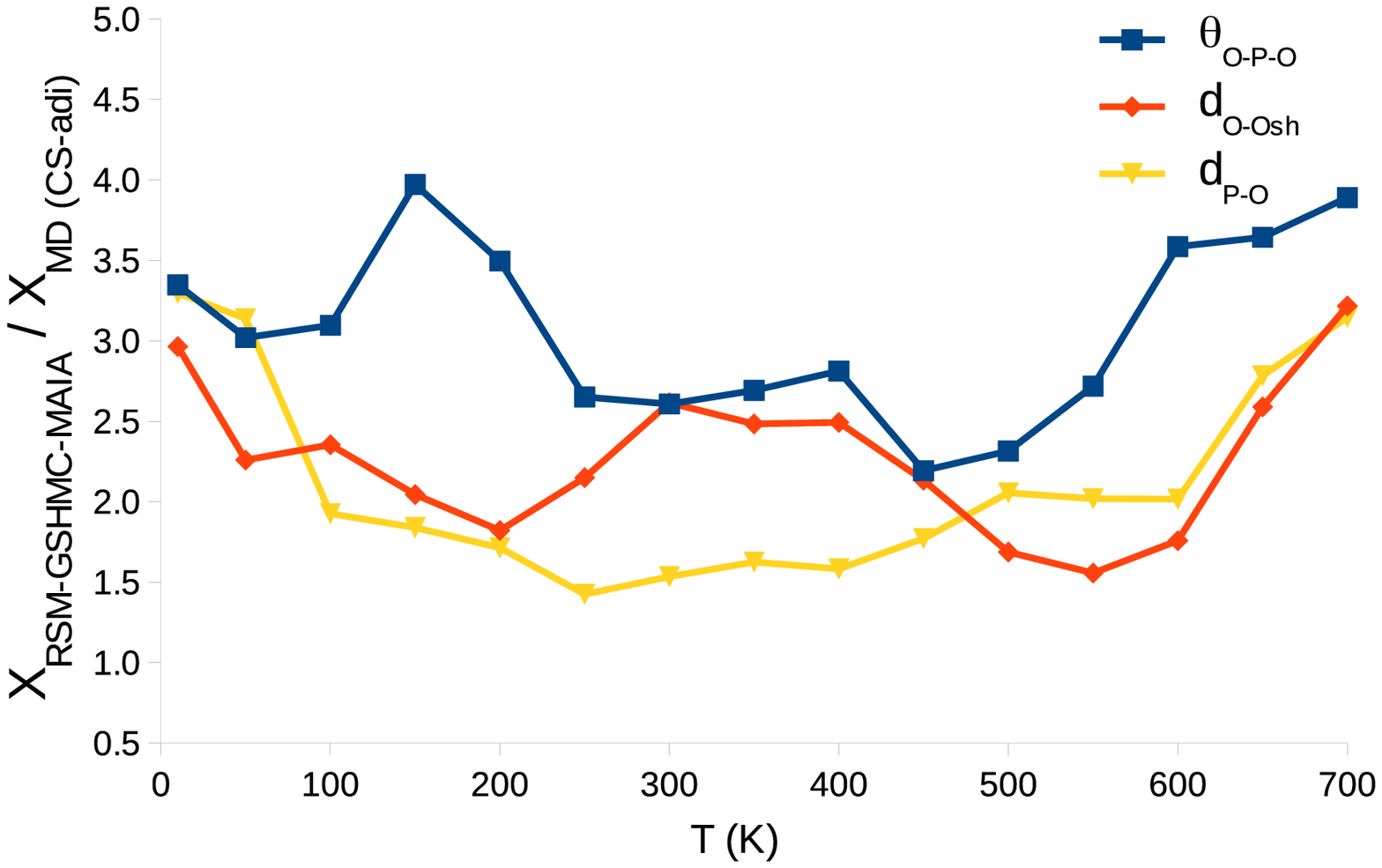}
\caption{Sampling performance ($X$) relative to MD at different temperatures achieved for several structural parameters when using the GSHMC (top) and RSM-GSHMC (bottom) methods.}
\label{fig:rel_perf_struct}       
\end{figure}

\begin{figure}
 \includegraphics[width=.5\textwidth]{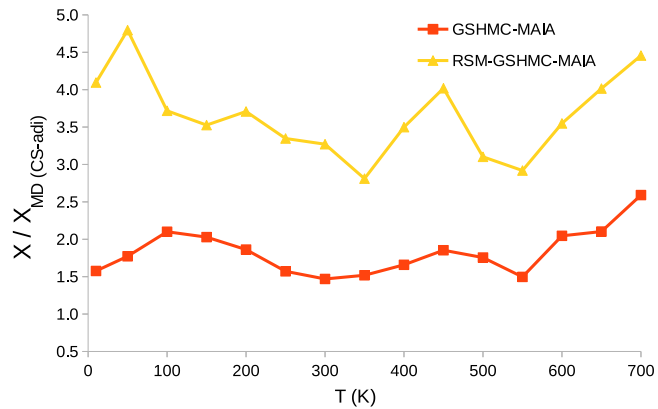}
\caption{Sampling performance ($X$) relative to MD at different temperatures achieved when computing the diffusion coefficients with GSHMC and RSM-GSHMC methods.}
\label{fig:rel_perf_diff}       
\end{figure}

In Figures \ref{fig:rel_perf_struct} and \ref{fig:rel_perf_diff} we present the performance of the methods for a range of temperatures and different quantities of interest in terms of relative $X$ values with respect to the obtained with MD. 
We can see that the GSHMC methods with and without mass randomization offer a significant improvement over MD (up to 2.5 and 4.7 for GSHMC and RSM-GSHMC respectively).

As we noticed in Figure \ref{fig:IACF_bars}, the RSM-GSHMC method is particularly beneficial for calculating diffusion coefficients. This is apparent at all temperatures (see Figure \ref{fig:rel_perf_diff}).
For other calculated properties RSM-GSHMC also demonstrates its superiority over MD and GSHMC at all temperatures though its performance differs less dramatically from the one offered by GSHMC.
Yet another advantage of RSM-GSHMC over other tested methods is that it can be further tuned by modifying the amount of randomized mass in  Eq. \eqref{eq:randomize} for each specific temperature. 

\section{Conclusions}
\label{sec:conclu}

We presented the new methodology for atomistic simulation of solid state materials for batteries, which offers a better accuracy and sampling efficiency than can be achieved with popular molecular dynamics (MD) approaches. 
The sampling in this method is performed with Generalized Shadow Hybrid Monte Carlo or GSHMC, which combines in a rigorous and effective manner molecular dynamics trajectories with Monte Carlo steps. 
The accuracy of the method is ensured by the new system specific adaptive integrators MAIA used in the MD step as well as by the modifications introduced in the adiabatic Core-Shell model for retaining the dynamics of a simulated system. 
Utilizing the adiabatic Core-Shell model or CS-adi in the new method instead of the Core-Shell relaxation scheme or CS-min, yields important performance gain saving up to 80\% of the computational time.
We have applied the method to the study of olivine NaFePO$_4$ systems and analyzed its accuracy and performance in comparison with available experimental data, the DFT computed properties, and the results obtained with other atomistic simulation methods (MD and conventional GSHMC).
The accuracy of the method in the calculation of lattice constants and thermal expansion has been compared against DFT-based calculations and experimental data, obtaining reliable results for all properties. 
Moreover, the method demonstrates a better agreement with the experimental data than one can observe with other tested atomistic methods, namely MD (CS-min), MD (CS-adi) and the original GSHMC.    

Introducing the novel MAIA integrator in our new methodology has also allowed for more efficient sampling when characterizing structural properties, such as average angles between atoms and bond lengths, as well as improving stability at higher temperatures.

Applying a randomization term to the shell mass improved not only the accuracy but also the sampling efficiency, especially when measuring diffusion coefficients. 
This modification of the GSHMC algorithm does not introduce significant overhead and is fully compatible with parallel implementations. 

In summary, the proposed methodology can be viewed as an alternative to molecular dynamics for atomistic studies of solid-state battery materials whenever high accuracy and efficient sampling are critical for obtaining tractable simulation results.

\begin{acknowledgements}
We acknowledge the financial support by grant MTM2013-46553-C3-1-P funded by MINECO (Spain). B.E. acknowledges the Iberdrola Foundation ``Grants for Research in Energy and Environment 2014''.
E.A. and T.R. thank for support Basque Government - ELKARTEK Programme, grant KK-2016/00026.
The SGI/IZO-SGIker UPV/EHU and the i2BASQUE academic network are acknowledged for computational resources.
M.F.P. would like to thank the Spanish Ministry of Economy and Competitiveness for funding through the fellowship BES-2014-068640.
This research was supported by the Basque Government through the BERC 2014-2017 program and by the Spanish Ministry of Economy and Competitiveness MINECO: BCAM Severo Ochoa accreditation SEV-2013-0323.
\end{acknowledgements}

\bibliographystyle{tca}

\end{document}